\documentclass{article}

\usepackage[sort&compress, numbers]{natbib}
\usepackage[final]{nips_2017}

\usepackage[utf8]{inputenc} 
\usepackage[T1]{fontenc}    
\usepackage{hyperref}       
\usepackage{url}            
\usepackage{booktabs}       
\usepackage{amsfonts}       
\usepackage{nicefrac}       
\usepackage{microtype}      
\usepackage{subfigure}
\usepackage{placeins}

\usepackage{times}
\usepackage{epsfig}
\usepackage{graphicx}
\usepackage{amsmath}
\usepackage{amssymb}
\usepackage{wrapfig,floatrow}

\usepackage{multirow}
\usepackage[lined,boxed,commentsnumbered,ruled]{algorithm2e}
\usepackage{adjustbox}
\usepackage{wrapfig}
\usepackage{caption}

\def\vec{\textsf{vec}} 

\newcommand{\eg}[0]{\emph{e.g. }}
\newcommand{\ie}[0]{\emph{i.e. }}

\newcommand{\RN}[1]{%
	\textup{\lowercase\expandafter{\it \romannumeral#1}}%
}

\newcommand{\Amat}[0]{{{\bf A}}}
\newcommand{\Bmat}{{\bf B}}
\newcommand{\Cmat}{{\bf C}}
\newcommand{\Dmat}{{\bf D}}
\newcommand{\Emat}[0]{{{\bf E}}}

\newcommand{\Gmat}{{\bf G}}

\newcommand{\Imat}{{\bf I}}

\newcommand{\Mmat}{{\bf M}}

\newcommand{\Rmat}[0]{{{\bf R}}}
\newcommand{\Smat}[0]{{{\bf S}}}

\newcommand{\Umat}[0]{{{\bf U}}}
\newcommand{\Vmat}[0]{{{\bf V}}}
\newcommand{\Wmat}[0]{{{\bf W}}}
\newcommand{\Xmat}[0]{{{\bf X}}}

\newcommand{\bv}[0]{{\boldsymbol{b}}}

\newcommand{\fv}[0]{{\boldsymbol{f}}\xspace}

\newcommand{\hv}[0]{{\boldsymbol{h}}}

\newcommand{\rv}{\boldsymbol{r}}
\newcommand{\sv}[0]{{\boldsymbol{s}}}

\newcommand{\yv}{\boldsymbol{y}}
\newcommand{\zv}{\boldsymbol{z}}

\newcommand{\Phimat}{\boldsymbol{\Phi}}

\newcommand{\alphav}{\boldsymbol{\alpha}}

\newcommand{\thetav}{\boldsymbol{\theta}}

\newcommand{\R}{\mathbb{R}}

\newcommand{\Ncal}{\mathcal{N}}

\newcommand{\Dcal}{\mathcal{D}}

\newcommand{\pre}{neural enhancing }
\newcommand{\ini}{seeds cleansing }
\newcommand{\Pre}{Neural Enhancing }
\newcommand{\Ini}{Seeds Cleansing }

\newcommand{\ctt}{\mathtt{c}}
\newcommand{\ftt}{\mathtt{f}}
\newcommand{\itt}{\mathtt{i}}
\newcommand{\ott}{\mathtt{o}}

\title{Seeds Cleansing CNMF for Spatiotemporal Neural Signals Extraction of Miniscope Imaging Data}
\author{Jinghao Lu$^1$, Chunyuan Li$^3$, Fan Wang$^{1, 2}$\\
	$^1$Department of Neurobiology and \\
	$^2$Cell Biology, Duke University Medical Center, and \\
	$^3$Electrical and Computer Engineering \\
	Duke University, Durham, NC 27708\\
	{\tt\small \{jinghao.lu, cl319, fan.wang\}@duke.edu}
}

\begin{document}	
	\maketitle
	
	\begin{abstract}

	Miniscope calcium imaging is increasingly being used to monitor large populations of neuronal activities in freely behaving animals. However, due to the high background and low signal-to-noise ratio of the single-photon based imaging used in this technique, extraction of neural signals from the large numbers of imaged cells automatically has remained challenging. Here we describe a highly accurate framework for automatically identifying activated neurons and extracting calcium signals from the miniscope imaging data, \ini Constrained Nonnegative Matrix Factorization (sc-CNMF). This sc-CNMF extends the conventional CNMF with two new modules: {\it i)} a \textit{neural enhancing} module to overcome miniscope-specific limitations, and {\it ii)} a \textit{seeds cleansing} module combining LSTM to rigorously select and cleanse the set of seeds for detecting regions-of-interest. Our sc-CNMF yields highly stable and superior performance in analyzing miniscope calcium imaging data compared to existing methods.

	\end{abstract}
	
	
	
	\section{Introduction}~\label{sec:introduction}
	\vspace{-6mm}

	In neuroscience, calcium imaging has become one of the staple technologies to simultaneously monitor the activities from a large population of neurons in awake and behaving animals~\cite{peron2015cellular,stirman2016wide,flusberg2008high,ghosh2011miniaturized}. 
	In most experiments, the target neurons express genetically encoded calcium indicators, and transiently change their fluorescence levels reflecting dynamic calcium concentrations within the neurons.	
	
	{\it Two-photon microscopy} is the most commonly used method for in vivo calcium imaging, which produces high quality videos over the imaging plane with relatively low and stable background~\cite{peron2015cellular,stirman2016wide}.
	A rich line of work has leveraged this advantage to extract calcium signals, including the popular PCA/ICA method~\cite{mukamel2009automated,reidl2007independent}, K-SVD dictionary learning \cite{pachitariu2013extracting}, Nonnegative Matrix Factorization \cite{maruyama2014detecting}, Constrained Nonnegative Matrix Factorization (CNMF)~\cite{pnevmatikakis2016simultaneous} and sliding window convolutional network~\cite{apthorpe2016automatic}. 
	However, animals are head-restrained and unable to move freely under the two-photon setup. 
	Furthermore, imaging deep brain regions in awake behaving animals using two-photon microscopy remains challenging.
	 
	Recently, advances in integrated {\it miniscope technology} enables imaging of large neural populations possible in freely behaving animals~\cite{cai2016shared,flusberg2008high,ghosh2011miniaturized}. 
	This technology allows neuroscientists to study cortical and subcortical neural circuits using a rich repertoire of animal behaviors. 
	However, there are two major challenges in adopting the miniscope imaging for neural activities with cellular resolutions recording. 
	First, the miniscope imaging is a single-photon-based technology that results in much noisier data, and innegligibly uneven and unstable background fluctuations, compared to two-photon microscopy~\cite{resendez2016visualization}. 
	These challenges result in the failure of directly applying previous two-photon imaging algorithms in properly extracting calcium signals from the miniscope imaging data.
	Second, neural activities during hundreds of behavioral sessions are often recorded, with each session easily containing GB size of data. 
	Therefore, manual annotation becomes increasingly laborious and even impossible in cases of large-scale datasets as well as unreliable (\textit{i.e.} failure in detecting overlapping and low-activity neurons). 
%
	\begin{figure}[t!]	
		\centering
		\subfigure[Raw Max Proj.]{%
			\includegraphics[width=0.25\textwidth]{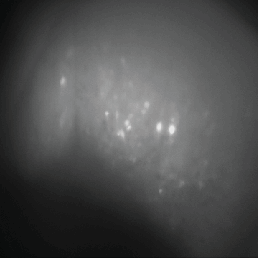} 
		} 
		\hfill
		\hspace{1mm}
		\subfigure[Processed Max Proj.]{%
			\includegraphics[width=0.25\textwidth]{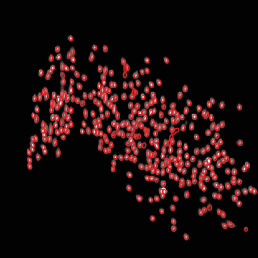} 
		} 
		\hfill
		\vspace{-2mm}
		\subfigure[Traces]{%
			\includegraphics[width=0.35\textwidth]{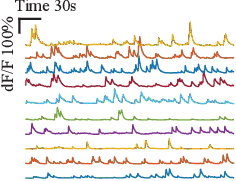} 
		}\\
		\subfigure[Pipeline]{%
			\includegraphics[width=1\textwidth]{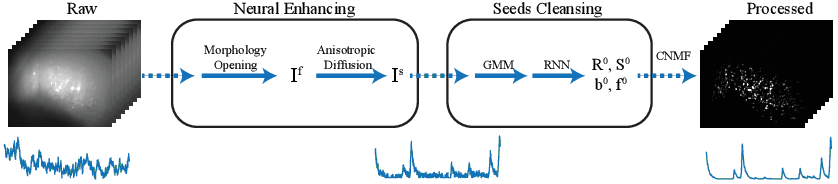} 
		} 
		\vspace{-2mm}
		\caption{Problem illustration: neural signals extraction from miniscope imaging. (a) Max projection of raw video. (b) Max projection of fully processed video with identified ROIs in the  contours. (c) Extracted neural signals. (d) Pipeline of the proposed methods. \label{fig:fig1}}
	\end{figure}

	To overcome these challenges, we develop the \ini Constrained Nonnegative Matrix Factorization (sc-CNMF) to extract calcium activity signals from the in vivo miniscope imaging videos. 
	Our sc-CNMF can accurately locate the regions of interest (ROIs), and yield the calcium signal of the corresponding neurons (see Fig.~\ref{fig:fig1}(a-c)). 
	Specifically, we extend the pure CNMF in two fronts, with the following contributions
	%
	%
	$\RN{1})$ A general {\it \pre}module is designed for the single-photon calcium imaging data that minimizes the influence of background instability and fluctuation. 
	$\RN{2})$ More importantly, the {\it \ini}module is proposed to reliably detect the population of ROIs, and thus significantly boost the performance in processing miniscope imaging data by maximizing both the precision and recall with increased efficiency.
%
	We emphasize that our \ini is non-trivial. 
	It consists of two steps: a Gaussian Mixture Model (GMM) followed by a Long Short-Term Memory (LSTM) based processing to identify seeds for ROIs~\cite{hochreiter1997long}. 
	A typical pipeline of our algorithm is shown in Fig.~\ref{fig:fig1}(d).	

	\begin{wraptable}{r}{8cm}\centering 
		\begin{minipage}{0.40\textwidth}
			\centering
			\caption{\small Methods Comparison. ``\checkmark'' indicates features integrated in the methods, while ``--'' indicates not.} \label{tab:ani}
			\vskip 0.0in
			\begin{adjustbox}{scale=0.8,tabular=l|ccc,center}
				\hline
				& {\bf sc-CNMF} & {\bf PCA/ICA} & {\bf CNMF} \\
				\hline 
				Neural Enhancing & \checkmark & Minor & --\\
				\hline 
				Seeds Cleansing & \checkmark & -- & Moderate\\ 
				\hline
				Spatial Localization & \checkmark & -- & \checkmark \\
				\hline
				Temporal Deconvolution & \checkmark & -- & \checkmark \\
				\hline
				Overlapping Separation & \checkmark & -- & \checkmark \\
				\hline
				Automatic Merging & \checkmark & -- & Moderate \\
				\hline
				Parameter Tuning & Minimal & \checkmark & \checkmark \\
				\hline 
			\end{adjustbox}
		\end{minipage}
	\end{wraptable}

	\paragraph{Related Work}
%
	Previous algorithms for automatic extraction of calcium signals from imaging videos can be divided into three categories: linear unsupervised basis learning methods~\cite{mukamel2009automated,reidl2007independent,pachitariu2013extracting}, nonlinear unsupervised basis learning methods~\cite{maruyama2014detecting,pnevmatikakis2016simultaneous}, and the supervised learning method~\cite{apthorpe2016automatic}.
	The first and most popular algorithm, PCA/ICA, applies PCA as a simple preprocessing step for dimension reduction and noise removal \cite{mitra1999analysis}, and then applies spatiotemporal-ICA to segment individual ROIs and their time series.
	Though proposed to be fast and efficient, algorithms such as PCA/ICA are endogenously based on the assumption of linearity, thus failing in separating spatially overlapping ROIs or returning localized ROIs. 
	Later works integrated more subtle considerations of nonlinearity~\cite{maruyama2014detecting,pnevmatikakis2016simultaneous}. 
	CNMF further combines nonlinearity in matrix factorization with simultaneous deconvolving spike trains from calcium dynamics~\cite{pnevmatikakis2016simultaneous}, which returns more spatially localized maps of ROIs compared to the previous methods. 
	A recent work, sliding window convolutional networks~\cite{apthorpe2016automatic}, is the first attempt to use supervised deep learning to tackle the problem. 
	However, these previous methods all depend on either sophisticated parameter tuning or only applicable for two-photon imaging. 
	In this work, we choose PCA/ICA and CNMF as our comparing references, and we summarize the features of the different algorithms in Table~\ref{tab:ani}.
%
%
%
	\section{Preliminaries on CNMF}
%
	The pure CNMF in our method is used to deconvolve neural populations and demix their calcium dynamics. 
	Specifically, for the video data $\Xmat \in \R^{P \times T}$ ($P$ is number of pixels per frame, and $T$ is number of frames), the CNMF decomposes $\Xmat$ into a spatial dictionary $\Rmat \in \R^{P \times K}$, potentially individual ROIs (where $K$ is the number of identified ROIs), and corresponding temporal dynamics matrix $\Smat \in \R^{T \times K}$, in addition to the background $\Bmat $  and the noise $\Emat$: $\Xmat = \Rmat\Smat^\top + \Bmat + \Emat$.
	Similarly, with the rank-1 assumption on $\Bmat$, CNMF decomposes the background $\Bmat = \bv \fv^\top$,  where $\bv \in \R^{P}$ and $\fv \in \R^{T}$.
%
	Meanwhile, $\Smat$ is also correlated with underlying action potential events: $\Amat = \Smat^\top \Gmat$, where $\Amat \in \R^{K \times T}$, and $\Gmat \in \R^{T \times T}$ is the second-order autoregressive matrix.
%
%
%
%
	The parameters  $\Rmat, \Smat, \bv, \fv$ in CNMF are estimated via iteratively alternating between the following two steps~\cite{pnevmatikakis2016simultaneous}.
	
	\paragraph{Estimating Spatial Parameters}
	Given the estimates of temporal parameters $\Smat^{(\ell-1)}$ and $\fv^{(\ell-1)}$ from the last iteration, the spatial parameters can be updated by solving the problem:
%
	$\min_{\Rmat, \bv} \|\Rmat\|_\mathrm{1}$, subject to $\Rmat, \bv \geq \mathrm{0}$ and $\|\Xmat(i, :) - \Rmat(i, :)\Smat^{(\ell-1)\top} - \bv(i) \fv^{(\ell-1)\top}\| \leq \epsilon_i\sqrt{T}$,
%
	where $\Xmat(i, :)$ is the $i$th row of the matrix $\Xmat$, $\bv(i, :)$ is the $i$th element of the vector. 
	$\epsilon_i\sqrt{T}$ is the empirically selected noise residual constraint of the corresponding pixel. This is essentially a basis pursuit denoising problem, and it is solved by the least angular regression~\cite{efron2004least} in implementation. 
	
	\paragraph{Estimating Temporal Parameters}
%
	Given the estimates of spatial parameters $\Rmat$, $\bv$ and temporal parameters  $\Smat^{(\ell-1)}$, $\fv^{(\ell-1)}$ from the last iteration, the temporal parameters can be updated by solving the problem:
	%
%
	$\min_{\Smat, \fv} \sum_{k=\mathrm{1}}^{K}\mathbf{1}^{\top} \Gmat\Smat$, subject to $\Gmat\sv_k \geq \mathrm{0}, k =1 \cdots K$ and $\|\Xmat(i,:) - \Rmat(i,:)\Smat^{(\ell-1)} - \bv(i)\fv^{(\ell-1)\top}\| \leq \epsilon_i \sqrt{T}$.

	\section{sc-CNMF}~\label{sec:methods}
	\vspace{-6mm}

	The peformance of pure CNMF is limited in the following three aspects:
	$\RN{1})$  
	It requires the product of an ROI guess and corresponding dynamics subtracted from the residual each iteration to initialize the rest ROI seeds. Therefore, the spatial maps of the remaining ROIs are dependent on the previous iterations, which renders CNMF very sensitive to the initial guesses of the ROI location.
	$\RN{2})$ 
	The number of ROIs $K$ is unknown in practice but the algorithm requires it manually set by users. CNMF will yield a higher true negative ratio if $K$ is too low, or many false positives if $K$ is set too high. While it is possible to manually decrease false positives, the missed ROIs can never be rescued back.
	$\RN{3})$ The iterative initialization nature of CNMF requires sequential execution of the algorithm.
	
	To overcome the above limitations, we developed two modules, the \pre and seeds cleansing, to provide good initialization for downstream modified pure CNMF in this section. 
	
	\subsection{\Pre}
	Existing works either employ na\"{\i}ve video cleaning operations (PCA/ICA) or assume low-rank background dynamics (rank-1 in CNMF). 
	However, in single-photon miniscope imaging, the background cannot be well approximated by rank-1 decomposition due to the non-uniform illumination and large temporal fluctuation of the background.
	Here we propose a two-step \pre module.
%
%
%
	\paragraph{Background Removal}
%
	We propose to separate the foreground neural signals from the background in a frame-wise fashion. 
	We denote $\Imat \in \R^{M \times J}$ as a single frame of the video $\Xmat$ before vectorization, \ie 
	$ \vec(\Imat) = \Xmat(:,t), t \in \{1, \cdots, T\}$, where $M, J$ are the width and height of the image.
	Based on the observation that neuronal ROIs are small in size compared to background structures, the gray-scale morphological opening operator with a flat {\it structuring element matrix} $\Phimat$ can well separate the background $\Bmat$ from the image $\Imat$ adaptively. 
	In particular, the opening operator is the combination of erosion $\ominus$ and dilation $\oplus$:
	$\Bmat = (\Imat \ominus \Phimat ) \oplus \Phimat$~\cite{van1992methods},
%
	where the morphological erosion returns minimum value
	%
	and dilation returns maximum value within the same structuring window at each point.	
%
%
%
	In pratice, the size of $\Phimat$ is set comparable to the size of ROIs in the imaging field. 
%
	The foreground is then computed as: $\Imat^{\mathtt{f}} = \Imat - \Bmat$.
	After the dynamic background removal, $\Imat^{\mathtt{f}}$ contains only neuronal information with minimal background corruption.	 

	

%

	
	\paragraph{Denoising}
%
	Though the influence of fluctuating background is maximally suppressed, the sparse yet sharp spatial noise randomly embedded within each frame remains possible to contaminate the neural information. To remove the spatial noise within ROIs and in the background, we apply denoising operation on $\Imat^{\mathtt{f}}$ using anisotropic diffusion~\cite{perona1990scale}. 
	For a given diffusion time $\tau$, the evolution follows the equation 
	$\frac{\partial \Imat^{\mathtt{f}}}{\partial \tau} = \mathrm{div} ( \Cmat  \nabla \Imat^{\mathtt{f}}) \triangleq \nabla \Cmat  \cdot \nabla \Imat^{\mathtt{f}} + \Cmat  \Delta \Imat^{\mathtt{f}}$
	where $\Cmat$ is the diffusive coefficient matrix depending on pixels and $\tau$. 
	By choosing concrete form of $\Cmat$ and $\tau$, we can control the smoothing level along or perpendicular to the boundaries between neurons and the background. 
	Due to the simple structures in the cleaned imaging field, we choose the classical Perona-Malik filter $\Cmat =\exp \left( -\frac{\| \nabla \Imat^{\mathtt{f}} \|^2}{\kappa^2} \right)$, where $\kappa$ controls the threshold of high-contrast.  
	The diffusivity is selected to preferentially smooth high-contrast regions, and $\kappa$ and $\tau$ are chosen to allow high tolerance.	
	The output image $\Imat^{\mathtt{s}}$ contains reduced spatial noise while preserves the boundary information of ROIs.
	

%

	
	\subsection{\Ini}
	%
	Previous work either contains no explicit initialization step to obtain the complete ROI set reliably (\textit{e.g.} PCA/ICA), or initializes by solving a convex optimization problem (CNMF). 
%
	However, these initializations usually fail in the case of highly dominant and dynamic background, where the approximation does not hold. 
	Moreover, a problem of previous initialization methods is that the subsequently initialized components are dependent on the preceding results, and the quality of the initialization significantly relies on the users' inputs of parameters. 
	We believe that, therefore, an accurate initialization process is necessary for achieving convergence to the correct ROI set as well as for reducing information loss/duplication.
	In this section, we generate and cleanse the potential seeds of ROIs, and create the initial spatial regions and the time series of these ROIs for the final refinement of spatiotemporal signals. 
	
	\paragraph{Over-complete Seeds Initialization}
%
	In order to generate the initial set containing centers of all potential ROIs, we construct a {\it randomized max pooling} process to create an over-complete set of seeds. 
	This process first randomly selects a portion of frames $\Imat^{\mathtt{s}}_{\alphav}= \{\Imat^{\mathtt{s}}_t\}_{t \in \alphav} , \alphav \subseteq \{1, \ldots, T\}$, where $\alphav$ is a randomized subset of frame number. 
	We then compute the max-projection map across selected frames $\Imat^{\text{max} } = \max(\Imat^{\mathtt{s}}_{\alphav} )$, and further detect all the local max points on this map as $\mathcal{S}$.
%
	We repeat the above procedures multiple times, and collect the union of each map $\mathcal{S}$ as the final over-complete set of neural seeds $\mathcal{S}_{\text{seeds} }$.  
	Our randomized max pooling can improve the true positive rate compared with max-pooling over the entire video, because a true seed can be buried in the uneven florescence of an ROI over a long period but can most likely be discovered in a small temporal vicinity.
	To exclude false positive seeds from the set, we propose the following two-stage algorithm for seeds refinement. 
	
	\paragraph{Seeds Refinement with GMM}
%
	The temporal properties of ROI and non-ROI can be very different.  
	ROIs often have prominent peak-valley difference $d$,  while non-ROIs tend to be less spiky.
	We assume $\{d_s | s\in \mathcal{S}_{\text{seeds}}\}$ are generated from a mixture of two Gaussians 
	$d_s \sim \sum_{i=1}^{2} \omega_i \Ncal(d | \mu_i,\sigma_i^2)$
	where $\mu_i,\sigma_i^2, \omega_i$ indicate the mean, variance, mixture proportion of the $i$th Gaussian component. 
	Therefore, we can cluster the seeds based on their probabilities of belonging to each component, and only consider those having higher probabilities to the Gaussian with a larger mean as positive seeds. 
	
	
		
	
	\paragraph{Seeds Refinement with LSTM}
	%
	To select true ROI seeds from the over-complete set, we need to characterize the patterns of neuronal calcium dynamics.
	Previous methods include using biophysical frameworks with minimal priors \cite{vogelstein2009spike}, and low-order autoregressive process \cite{vogelstein2010fast}, with the assumption that there exists clear stereotypical calcium spike patterns. However, in the miniscope imaging data intrinsically corrupted with large noise, the sequences (the neuronal calcium trace) can be complex. Hence we propose to employ Recurrent Neural Networks (RNNs)~\cite{lecun2015deep} for calcium signal sequence modeling and classification. 
	We offline train the RNNs with a separate training dataset, composed of both positive and negative sequence chunks of length $T_0 = 100$, obtained with $10$Hz frame rate. 
	The positive sequences are real {\it calcium spikes} cropped from the training video aligned to the peak of the spike, whereas the negative sequences are randomly selected time periods from the rest non-spike periods. 
	Specifically, consider training data $\Dcal = \{\Dmat_1, \cdots, \Dmat_N\}$, where $\Dmat_n \triangleq (\yv_n, l_n) $, with input sequence $\yv_n$ and output label $l_n \in \{0,1\}$.
	Our goal is to learn model parameters $\thetav$ to best characterize the mapping from $\yv_n$ to $l_n$ with likelihood $p(\Dcal | \thetav) = \prod_{n=1}^N p(\Dmat_n | \thetav)$. In our setting for sequence classification, the input is a sequence, $\yv=\{y_1,\ldots,y_{T_0}\}$, where $y_t$ is the input data at time $t$. There is a corresponding hidden state vector $\hv_t \in \R^{K}$ at each time $t$, obtained by recursively applying the {\em transition function} $\hv_t = g( \hv_{t-1}, \yv_t; \Wmat, \Umat)$. $\Wmat$ is \emph{encoding weights}, and $\Umat$ is \emph{recurrent weights}.
	The ouput $c$ for our classification is defined as the corresponding {\em decoding function} $p(c |\hv_{{T}_0}; \Vmat ) = \sigma (\Vmat \hv_{{T}_0}) $, where $\sigma(\cdot)$ denotes the {\it logistic sigmoid function}, and $\Vmat$ is \emph{decoding weights}.
	
	The transition function $g (\cdot)$ can be implemented with a \emph{gated} activation function, such as LSTM~\cite{hochreiter1997long} or a Gated Recurrent Unit (GRU)~\cite{cho2014learning}. Both LSTM and GRU have been proposed to address the issue of learning long-term sequential dependencies. 
	Each LSTM unit has a cell containing a state $\ctt_t$ at time $t$. This cell can be viewed as a memory unit. Reading or writing the memory unit is controlled through sigmoid gates: input gate $\itt_t$, forget gate $\ftt_t$, and output gate $\ott_t$. The hidden units $\hv_t$ are updated as follows:
	\begin{equation}\label{eq:lstm}
	\begin{split}
	\hspace{-4mm}
		\itt_t &= \sigma (\Wmat_{i}\yv_t + \Umat_{i}\hv_{t-1} + \bv_i)\,,\\	
		\ftt_t &= \sigma (\Wmat_{f}\yv_t + \Umat_{f}\hv_{t-1} + \bv_f)\,,\\	
		\ott_t &= \sigma (\Wmat_{o}\yv_t + \Umat_{o}\hv_{t-1} + \bv_o)\,,\\		
	\end{split}
	\quad\quad
	\begin{split}
		\tilde{\ctt}_t &= \tanh (\Wmat_{c}\yv_t + \Umat_{c}\hv_{t-1} + \bv_c)\,,\\
		\ctt_t &= \ftt_t \odot \ctt_{t-1} + \itt_t \odot \tilde{\ctt}_t\,,\\
		\hv_t &= \ott_t \odot \tanh(\ctt_t)\,,\\
	\end{split}
	\end{equation}
	%
	%
	where $\odot$ represents the element-wise matrix multiplication operator. Note that the training of RNNs is completed off-line, only the efficient testing stage is performed for seeds refinement.
		
	In the testing stage, given an input $\tilde{\yv}$ (with missing label $\tilde{l}$), the estimate for the output is $ p( \tilde{l} | \tilde{\yv}, \hat{\thetav})$, where $\hat{\thetav} = \arg \max \log p( \Dcal | \thetav ) $. 
	In our practical application, the testing sequence $\bar{\yv} \in \R^{T}$ is often of length $T > T_0$. 
	We first convert it into a bag of subsequences $\{ \tilde{\yv}_i \}_{i=1}^{T-T_0+1}$,  with a sliding window of width $T_0$ and moving step size $1$. 
	The well trained LSTMs are then used to label the subsequences. 
	We consider $\bar{\yv}$ as positive if at least one subsequence in its bag is classified as "calcium spike" $(l=1)$, otherwise negative. Intuitively, this means that we only care about the patterns of neural spikes, regardless of their temporal positions. We will show the performance of the classifier in the Experiment section.

	\paragraph{Seeds Merging}
	Once the set of seeds is cleansed, there is still a low possibility of identifying multiple seeds within a single ROI. 
	Therefore, we merge all these redundant seeds by computing the temporal similarity of seeds within their neighborhood, and preserving the one with maximum intensity. 
	Specifically, we compute the similarity based on phase-locking information~\cite{hahn2006phase}.
	For a sequence, we extract the instantaneous phase dynamics using Hilbert transform, and only consider the subsequences containing prominent peaks, because all the pixels within the same ROI are highly correlated only during the calcium spiking periods, but not necessarily during baseline period.
	After seeds merging, we obtain $K$ seeds, as the number of ROIs in our sc-CNMF.

	%
	
	%
%
	
%
	
	\paragraph{Spatial and Temporal Initialization}
	The time series of the $k$th seed is used as initial guess of temporal signal $\hat{\sv}_k$. 
	The spatial map of the corresponding ROI, $\hat{\rv}_k$, is estimated by pooling neighbor pixels with temporal similarity above a threshold. 
	Because of our seeds generating and cleansing approach, the spatiotemporal initialization in our method can be readily parallelized.
	Then we fine-tune the corresponding spatial and temporal guess, by applying semi-NMF:
	$\min \| \zv_k - \hat{\rv}_k \hat{\sv}_k^\top \|$,
	where $\zv_k \subset \Imat^{\mathtt{s}}$ is a region containing the $k$th ROI. The spatial map is updated as
	$\rv_k \gets \zv_k  \sv_k (\sv_k^\top \sv_k)^{-1}$,
	and the temporal signal is updated as
	$\sv_k \gets \sv_k \sqrt{\frac{(\rv_k \zv_k)^+ + \sv_k (\rv_k^\top \rv_k)^-}{(\rv_k \zv_k)^- + \sv_k (\rv_k ^\top \rv_k)^+}}$,
%
	where $\Mmat^+ \triangleq \frac{|\Mmat| + \Mmat}{2} $ and $\Mmat^- \triangleq \frac{|\Mmat| - \Mmat}{2} $ for any matrix $\Mmat$~\cite{ding2010convex}.
	In total, we now have $K$ ROIs $\Smat^0 = [\sv_1,\cdots, \sv_K]$, and temporal signals $\Rmat^0 = [\rv_1, \cdots, \rv_K]$. 
	Similarly, the background parameters $\bv^0, \fv^0$ are also estimated by the same semi-NMF procedure, using $(\Imat^{\mathtt{s}} - \sum_{k=1}^{K} \rv_k \sv_k^\top)$. 
	$\Smat^0, \Rmat^0, \bv^0, \fv^0$ are then used as initializations of CNMF.

%
%
%

	
	\begin{figure}[t!]
		\centering		
		\subfigure[Identified ROI Contours]{
			\includegraphics[width=0.32\textwidth]{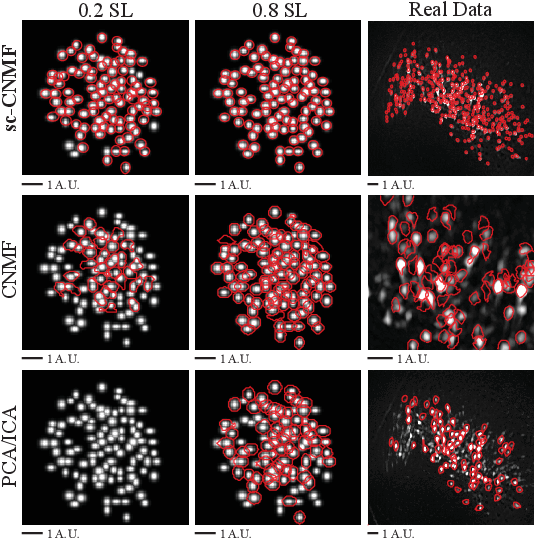} 
		}
		\hfill
		\subfigure[Spatiotemporal Correlation]{	 
			\includegraphics[width=0.64\textwidth]{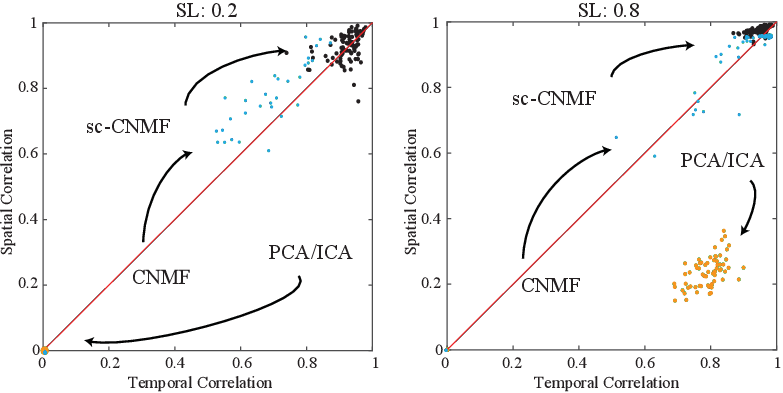} 
		}
		\vspace{-2mm}
		\caption{Visualization of the general performance. (a) A comparison of identified ROIs with contours between sc-CNMF, CNMF and PCA/ICA in simulated data of signal level (SL) $0.2 ~\& ~0.8$ and real data. (b) Spatiotemporal correlation between the identified ROIs and ground truth by the three methods in simulated data of SL $0.2 ~\& ~0.8$.}\label{fig:fig3}
	\end{figure}	
\section{Experiments}~\label{sec:experiments}
\vspace{-6mm}

We demonstrate the performance of sc-CNMF on both synthetic and real data. For simulation, we employ a scoring criterion that evaluates the performance, where we calculate the spatial and temporal similarity between the ground truth and the identified ROIs using rescaled cosine similarity. 
To measure the precision of the performance, we calculate true positive rate, false positive rate and false negative rate of the identified ROIs.
This can reflect the level of spatial locality objectively. For real data, we use the contours to annotate the identified ROIs, and superimpose them to max projection of preprocessed videos to show the performance. The proposed methods are implemented in Matlab.
We will release the codes once the paper is published. We also provide full resolution figures of visualization as well as an ablation study of each step in \hyperref[supp]{supplemental materials}.

\vspace{-2mm}
\paragraph{Simulation}
%
We synthesized 100 neurons with average neuron diameter $\gamma_0$ comparable to real data in a $128 \times 128$ field of view. 
Each neuron was simulated by a 2D Gaussian function with a variance in its two dimensions ($\gamma = \gamma_0 + \xi, ~\xi \sim \Ncal(0,1)$), and the calcium dynamics was generated by the first-order autoregressive process (decay time constant 0.95) with randomized spike events (spiking probability equals $1\%$), modified from \cite{pnevmatikakis2016simultaneous}. 
The background was extracted from the real single-photon miniscope imaging videos \cite{zhou2016efficient}, and temporal fluctuations of the background and spatial noises were added. 
The ground truth neural signals were then synthesized with the noisy background with a varying ratio, {\it signal level} (SL), ranging from $0.05$ to $0.8$ with a step size $0.05$. 
Due to the biophysical property, the fluorescence is positively correlated to the light stimulation, thus the spatial positions of simulated neurons were densely generated in the high background illuminated area.
%
\paragraph{Real Data}
%
Miniscope calcium imaging data were collected from the barrel cortex of awake and freely behaving mice.
The size of the raw videos were $1080 \times 1440 \times T_0$, with sampling rate $20$Hz.
For the performance test, we spatially downsampled the videos with a factor of $2$ ($540 \times 720$), and the temporal downsampling rate was $2$.
All experiments were conducted according to protocols approved by the Duke University Institutional Animal Care and Use Committee.
	
	\begin{figure}[t!]
		\centering
		\subfigure{ 
			\includegraphics[width=0.28\textwidth]{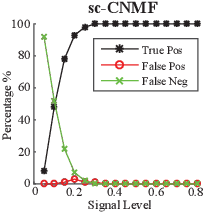} 
		}
		\hfill
		\addtocounter{subfigure}{-1}
		\subfigure[Identified ROI Summary]{ 
			\includegraphics[width=0.28\textwidth]{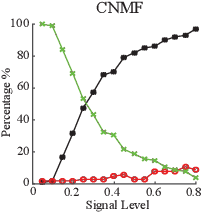} 
		}
		\hfill
		\addtocounter{subfigure}{-1}
		\subfigure{ 
			\includegraphics[width=0.28\textwidth]{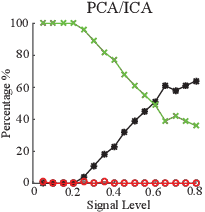} 
		}\\
		\vspace{-2mm}
		\centering		
		\subfigure[Mean Spatiotemporal Correlation]{ 
			\includegraphics[width=0.48\textwidth]{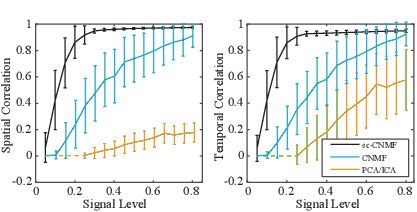} 
		}  
		\hfill
		\subfigure[Performance of the LSTM classifier]{	 
			\includegraphics[width=0.48\textwidth]{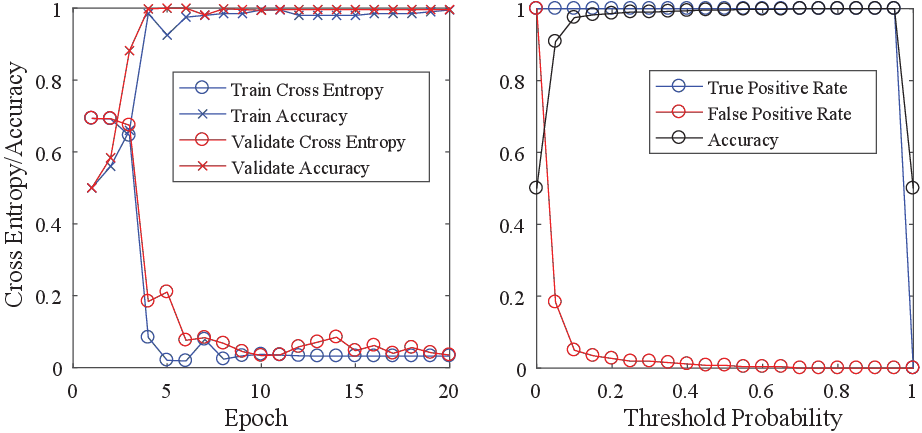} 
		}  
		\vspace{-2mm}
		\caption{Quantification and comparison of the general performance. (a) A comparison of the average spatiotemporal correlation between the identified ROIs and the ground truth by the three methods. (b) Accuracy and precision of the three methods. (c) Performance of the LSTM classifier.}\label{fig:fig4}
	\end{figure}
\vspace{-2mm}
	\subsection{General Performance}
	We compare our methods with PCA/ICA and pure CNMF. PCA/ICA is the most popular semi-automatic method for processing miniscope calcium imaging \cite{resendez2016visualization}. 
	We use the commercially available $\mathtt{Mosaic}$ software (Inscopix Inc.) \cite{jennings2015visualizing} as PCA/ICA implementation, and process the data by following the standard work-flow in the software manual. In particular, we choose the number of principal components (PC) and independent components (IC) based on their sugguested rate (\eg $20\%$ more ICs and $50\%$ more PCs than the estimated number of ROIs). For CNMF, we use the default initialization strategy as described in ~\cite{pnevmatikakis2016simultaneous}.
	
	In Fig.~\ref{fig:fig3}, we visualize the ROI detection results in (a) and their spatialtemporal correlation in (b).
	For both the real data and the simulations with signal levels at either $0.2$ or $0.8$, sc-CNMF can label a nearly complete set of ROIs, whereas the other methods only label a small subset of ROIs, and perform poorly in extracting correct ROIs.
	The spatial and temporal correlation between ground truth and identified ROIs are significantly higher with the use of sc-CNMF than with the PCA/ICA method.
	
	
	We show the performance comparisons for the full range of signal level in Fig.~\ref{fig:fig4}. 
	The overall general performance of all methods decreases, as signal level diminishes. However, the proposed sc-CNMF remains highly effective within a reasonably wide signal range, and significantly outperform PCA/ICA across all SLs. Compared to the pure CNMF, sc-CNMF performs evidently superior at low and intermediate SLs, and more stably with optimal results at high SLs. Notice that the constantly low spatial correlation of PCA/ICA is due to its lack of spatial localization of identified ROIs.

	\subsection{Performance of Each Module}
\begin{figure}[t!]
		\centering		
		\subfigure[Identified ROI Contours]{ 
			\includegraphics[width=0.38\textwidth]{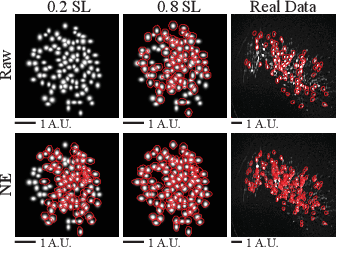} 
		}
		\hfill
		\subfigure[Mean Spatiotemporal Corr.]{	 
			\includegraphics[width=0.27\textwidth]{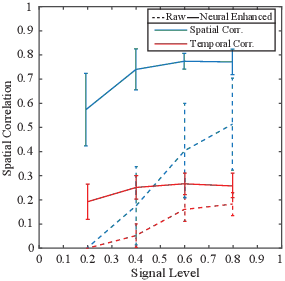} 
		}
		\hfill
		\subfigure[Identified ROI Summary]{ 
			\includegraphics[width=0.27\textwidth]{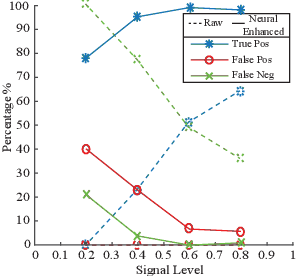} 
		}
		\vspace{-2mm}
		\caption{Improvements of using the neural enhanced (NE) video in PCA/ICA. (a) Visualization of the identified ROIs with contours using the raw or the NE video as an input to PCA/ICA. (b) A comparison of the average spatiotemporal correlation between the identified ROIs and the ground truth using the raw or the NE video. (c) Accuracy and precision of the two inputs.}\label{fig:fig5}
	\end{figure}

	\paragraph{\Pre}
	%
	We examine the effectiveness of the \pre module, in terms of eliminating the miniscope-specific background effect. 
	This is validated by comparing the performance of PCA/ICA method with and without \pre module. 
	For simulated data, we run the experiment on signal level $0.2$ to $0.8$ with a step of every $0.2$, with the same PC and IC rate (here PC $=150$ and IC $=120$). 
	The results are shown in Fig.~\ref{fig:fig5}. 
	We find that our neural enhancing module significantly improves the performance of PCA/ICA. The number of identified ROIs and overall spatiotemporal correlation significantly increase at all SLs.
	Similarly, in the case of real data, the number of identified ROI increases compared to using raw data. 
	There are still potential false positive and duplicated ROIs, but this can only be possibly overcome by manual intervention, which is due to the limitations of PCA/ICA method itself.

\begin{figure}[t!]
		\centering		
		\subfigure[Identified ROI Contours]{ 
			\includegraphics[width=0.38\textwidth]{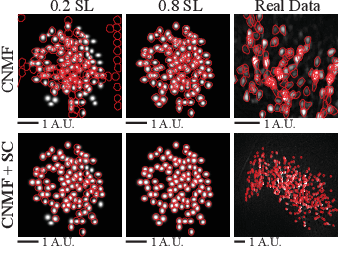} 
		}
		\hfill
		\subfigure[Mean Spatiotemporal Corr.]{	 
			\includegraphics[width=0.26\textwidth]{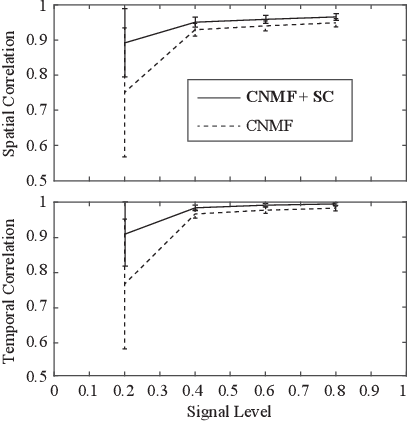} 
		}
		\hfill
		\subfigure[Identified ROI Summary]{
			\hspace{-1mm}	 
			\includegraphics[width=0.27\textwidth]{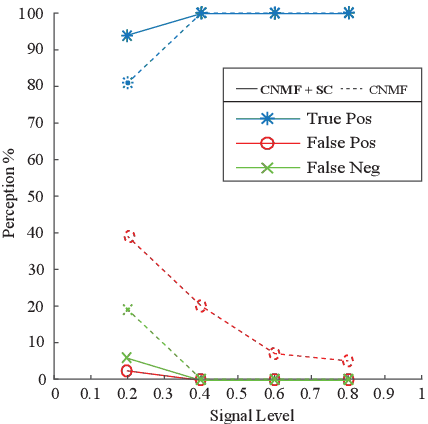} 
		}
		\vspace{-2mm}
		\caption{Improvements of the seeds cleansing (SC) in CNMF. (a) Visualization of the identified ROIs using the pure CNMF or the SC. (b) A comparison of the average spatiotemporal correlation between the identified ROIs and the ground truth with the pure CNMF or the SC. (c) Accuracy and precision of the two processes.}\label{fig:fig6}
	\end{figure}

	\paragraph{\Ini}
	We examine the importance of the \ini module in accurately selecting the valid set of ROI seeds. 
	This is implemented by replacing the initialization module in the pure CNMF with our seeds cleansing module. 
	The results are shown in Fig.~\ref{fig:fig6}. 
	In the default initialization of CNMF, we chose $K = 120$ to provide the tolerance of false positives. 
	Despite this, there exist several neurons not detected. 
	Because $K$ is greater than the ground truth, we see that the pure CNMF can (1) miss true ROIs, (2) generate duplicate seeds within a single ROI, and (3) include false positives when signal-to-noise ratio is not high enough.
	In the initialization of our sc-CNMF, a maximally correct set of ROIs is returned, without tuning the number of ROIs by trial-and-error.
	The last but not the least note is that for real data, CNMF cannot successfully finish computation of the full size dataset, due to memory overflow with the same hardware setting, thus we can only crop the video to compute a small patch.
	sc-CNMF, on the other hand, does not have this problem. 
	We provide a demo video as a supplement, showing the raw, neural enhanced and fully processed results by sc-CNMF from real data (\href{https://goo.gl/XnTw3C}{https://goo.gl/XnTw3C}).

	\section{Conclusions}~\label{sec:conclusion}
	\vspace{-6mm}
%

	In this work, we develop the \pre and \ini modules to provide high quality initialization to the modified CNMF framework. 
	These two novel modules, integrated with the modified CNMF framework, allow us to accurately extract neural signals from miniscope calcium imaging. 
	For neural enhancing, we employ frame-wise morphological opening to approximate the dynamic background in imaging videos, and to smooth the signals by anisotropic diffusion. 
	For seeds cleansing, we used GMM as a coarse seeds classifier, and a custom RNN/LSTM to detect a tight set of valid seeds. 
	Notice that the seeds cleansing module is the first method to integrate deep learning in the field of neural signal extraction of single-photon based calcium imaging.
	We post-process the set of seeds by merging and ``same-ROI suppressing'' to form an accurate final set of ROI seeds. 
	We then integrated the modified CNMF framework for refining spatial maps and temporal signals of detected ROIs. 
	The experimental results on both synthetic and real data confirm that sc-CNMF outperforms the PCA/ICA and pure CNMF methods, and is the first seed-accurate framework in extracting spatiotemporal calcium signals from miniscope based single-photon imaging.

\newpage
{\small
	\setlength{\bibsep}{0.98pt}
	\bibliographystyle{unsrtnat}
	\bibliography{sc-CNMF_ref}
}

\newpage
\noindent\makebox[\linewidth]{\rule{\linewidth}{3.5pt}}
\begin{center}
	\bf{\Large Supplementary Material of Seeds Cleansing CNMF}
\end{center}
\noindent\makebox[\linewidth]{\rule{\linewidth}{1pt}}

\appendix

\section{Additional Results}\label{supp}
\subsection{Full Scale Visualizations of the Imaging Data}
For better visualization of the identified ROI contours, we provide the larger-scale figures. The overall performance of the proposed sc-CNMF is in Fig.~\ref{fig:figs1_scCNMF}, the improvement by neural enhancement module is in Fig.~\ref{fig:figs2_NE}, and the improvment by seeds cleansing module in Fig.~\ref{fig:figs3_SC}.

For further investigation of the effectiveness of each module, we did the ablation study on each module, and  visualize the results as the max projection in Fig.~\ref{fig:figs4} \& Fig.~\ref{fig:figs5}, and the detected seeds in Fig.~\ref{fig:figs6}.

\FloatBarrier
\subsection{Notes on the Supplemental Video}
The {\it Raw} video shows the raw imaging data, while the {\it Preprocessed} and the {\it Processed} video represents the video after neural enhancing and the full sc-CNMF separately. Specifically, after \pre we can get the preprocessed video with every frame enhanced, but no separated spatiotemporal information of individual ROIs. However, after the whole process we do get every ROI and its corresponding temporal signal separated. We then take the product of the separated spatial and temporal matrix to retrieve the video version as the processed video. We also include the video in the supplemental materials.

\renewcommand{\thefigure}{S\arabic{figure}}

\setcounter{figure}{0}
	
	\begin{figure}[!h]
		\centering 
		\includegraphics[width=1\textwidth]{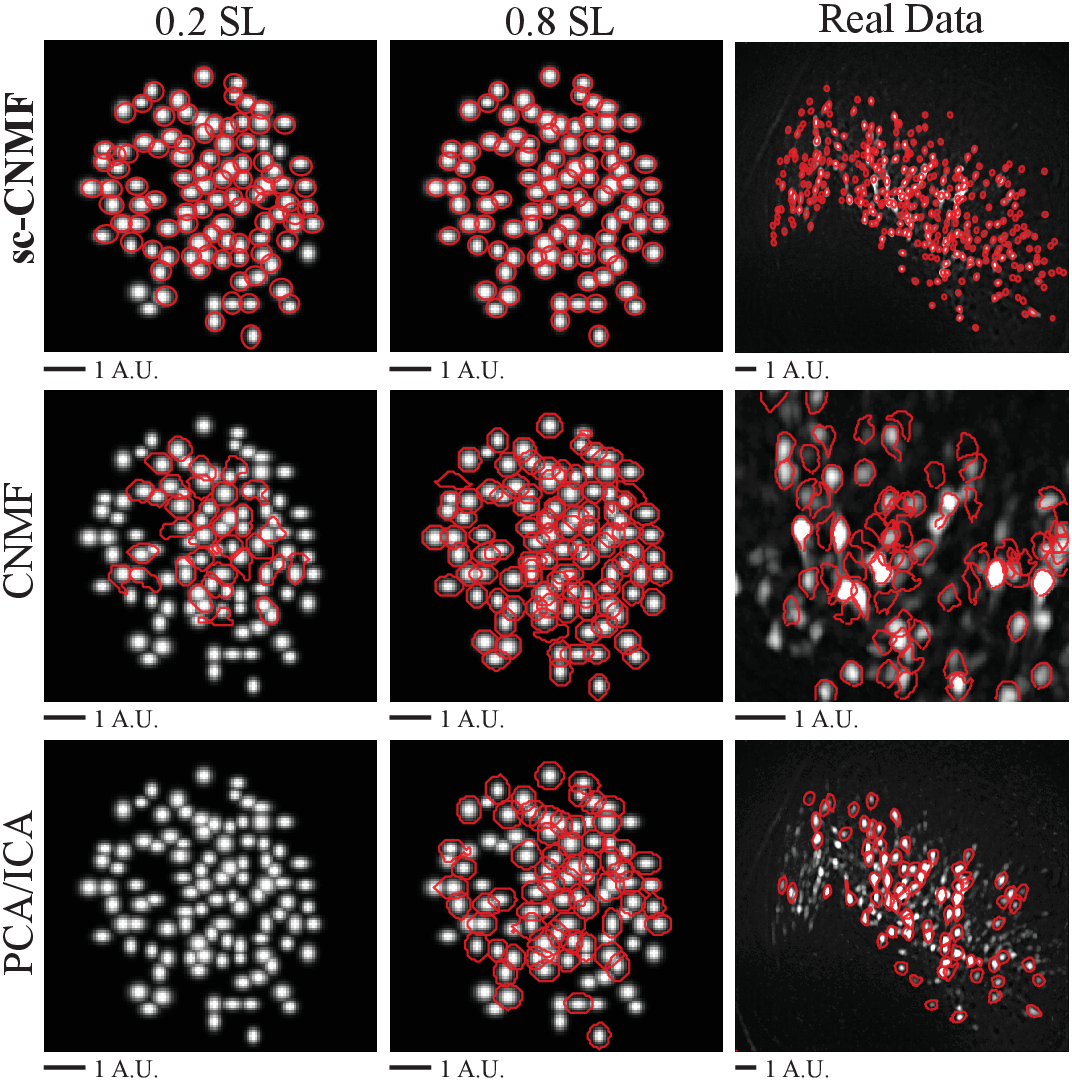} 
		\vspace{-5mm}
		\caption{Visualizations of the identified ROI contours of the three methods on simulated datasets (SL 0.2 \& 0.8) and real data.}\label{fig:figs1_scCNMF}
	\end{figure}
	
	\begin{figure}
		\centering 
		\includegraphics[width=1\textwidth]{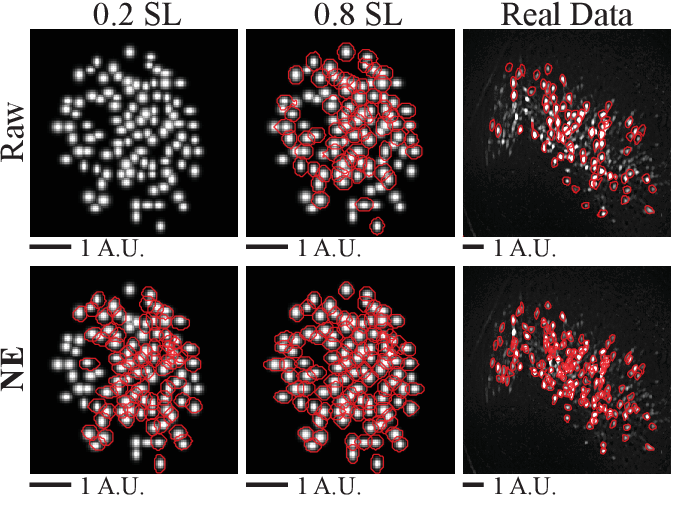} 
		\vspace{-5mm}
		\caption{Visualizations of the identified ROI contours using raw data or NE data as input, in the cases of simulated datasets (SL 0.2 \& 0.8) and real data.}\label{fig:figs2_NE}
		\vspace{-5mm}
	\end{figure}	
		\vspace{-5mm}
	\begin{figure}
		\centering
		\includegraphics[width=1\textwidth]{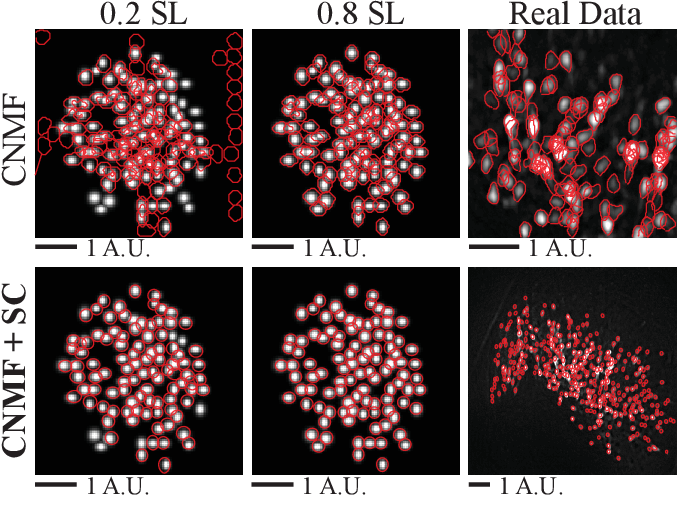} 
		\vspace{-10mm}
		\caption{Visualizations of the identified ROI contours using the pure CNMF or the seeds cleansing module, in the cases of simulated datasets (SL 0.2 \& 0.8) and real data.}\label{fig:figs3_SC}
	\end{figure}

	\FloatBarrier
	\newpage
	\subsection{Detailed Visualization of the Effect at Each Step}
		\vspace{-3mm}
	\begin{figure}[!h]
		\centering			
		\subfigure{ 
			\includegraphics[width=0.09\textwidth]{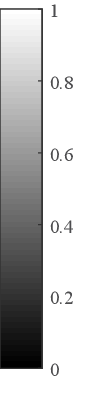} 
		}
		\hfill
		\addtocounter{subfigure}{-1}
		\subfigure[Max Proj. of Raw Data]{	 
			\includegraphics[width=0.4\textwidth]{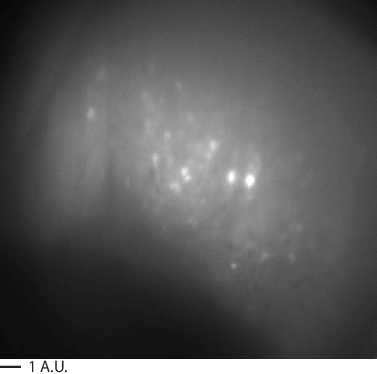} 
		}
		\hfill
		\subfigure[Max Proj. of BG Removed Data]{	 
			\includegraphics[width=0.4\textwidth]{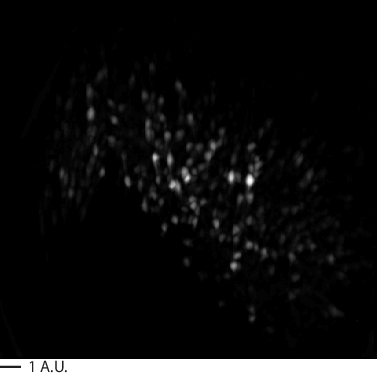} 
		}
		\vspace{-3mm}
		\caption{Visualizing the effect of background (BG) removal. (a) Max projection of the raw data, which contains significantly more BG noise as well as unbalanced and dominating BG illumination, compared to (b) the max projection of the BG removed version, which optimally removes BG illumination and maximally suppresses noises.}\label{fig:figs4}
	\end{figure}
		\vspace{3mm}
	\begin{figure}[!h]
		\centering			
		\subfigure{ 
			\includegraphics[width=0.09\textwidth]{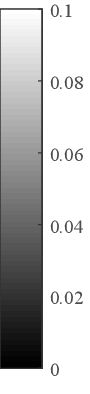} 
		}
		\hfill
		\addtocounter{subfigure}{-1}
		\subfigure[Max Proj. (Unsmoothed \& BG Removed)]{	 
			\includegraphics[width=0.4\textwidth]{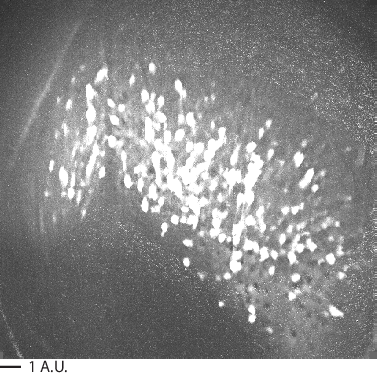} 
		}
		\hfill
		\subfigure[Max Proj. (Smoothed \& BG Removed)]{	 
			\includegraphics[width=0.4\textwidth]{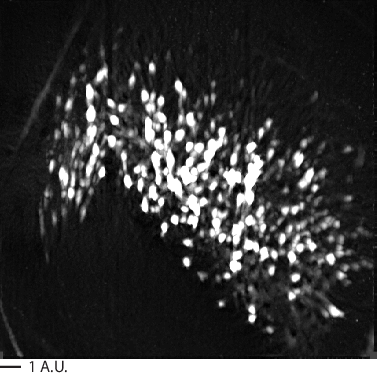} 
		}
		\vspace{-3mm}
		\caption{Visualizing the effect of smoothing. (a) Max projection of the background (BG) removed data after smoothing, which results in a much higher signal-to-noise Ratio, compared to (b) the max projection of the unsmoothed version. The intensity range of all the pixel is normalized to $[0, 1]$)}\label{fig:figs5}
	\end{figure}
	\vspace{-35mm}
	\begin{figure}[!h]
		\centering			
			\subfigure{ 
				\includegraphics[width=0.11\textwidth]{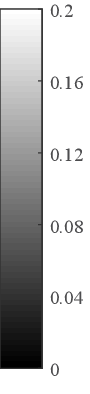} 
			}		
		\addtocounter{subfigure}{-1}
		\subfigure[Over-complete Seeds]{	 \centering			
			\includegraphics[width=0.48\textwidth]{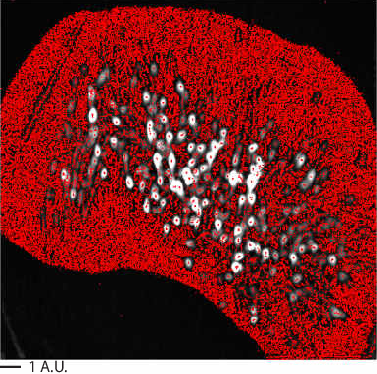} 
		}  \\
		\subfigure[Seeds Cleansed after GMM]{	 \centering			
			\includegraphics[width=0.48\textwidth]{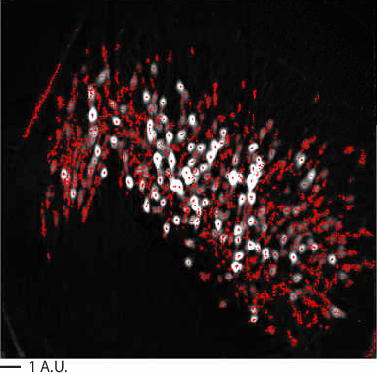} 
		} 
		\subfigure[Seeds Cleansed after LSTM]{	 \centering			
			\includegraphics[width=0.48\textwidth]{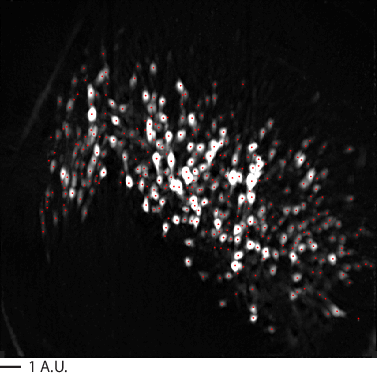} 
		}

			\vspace{-2mm}
		\caption{Visualizing the effect of GMM \& LSTM. (a) Over-complete set of seeds found as local maxima. (b) Seeds cleansed using GMM. False positives are significantly removed while true positives are remained. (c) Seeds cleansed using LSTM. False positives are maximally eliminated while only one seed per ROI is remained.}\label{fig:figs6}
	\end{figure}

\end{document}